\documentclass{article}

\usepackage{geometry}
\usepackage{mathpazo}
\usepackage{amssymb}
\usepackage[fleqn]{amsmath}
\usepackage{graphicx}
\usepackage{epstopdf}
\usepackage[titletoc,title]{appendix}
\usepackage{gensymb}
\usepackage{bm}
\usepackage{authblk}
\usepackage{xcolor}

\title{Non-Resonant Picosecond Three-Wave Mixing in the Gas Phase}

\author[1,$^\dagger$]{Grayson LaCombe}
\author[1,$^\dagger$]{Jianan Wang}
\author[2]{J\'er\'emy R. Rouxel}
\author[1,*]{Marien Simeni Simeni}

\affil[1]{Department of Mechanical Engineering, University of Minnesota, 111 Church St SE, Minneapolis, MN, 55455, United States of America}
\affil[2]{Chemical Sciences and Engineering 354 Division, Argonne National Laboratory, \\
Lemont, Illinois 355 60439, United States}

\affil[*]{msimenis@umn.edu}
\affil[$^\dagger$]{These authors contributed equally}

\begin{document}

\maketitle

\begin{abstract}
We report on the experimental observation of non-resonant, second-order optical Sum-Frequency Generation (SFG) in five different atomic and molecular gases. The measured signal is attributed to a SFG process by characterizing its intensity scaling and its polarization behavior. We show that the electric quadrupole mechanism cannot explain the observed trends. Our results demonstrate that SFG in the gas phase is about four orders of magnitude stronger than Third Harmonic Generation (THG) and independent from any externally-applied electric fields. These features make this method suitable for gas number density measurements at the picosecond timescale in reactive flows and plasmas.
\end{abstract}

\section{Introduction}
A classic textbook demonstration \cite{shen1984principles, boyd2008nonlinear, reintjes2012nonlinear} shows that even-order nonlinear susceptibilities vanish in centrosymmetric media.
It stems from the sign flip of electric dipole moments upon an inversion operation.
This property has made even-order techniques useful for probing interfaces \cite{miranda1999liquid, shen1986surface, geiger2009second} or chiral liquids \cite{belkin2001sum, fischer2000three}.
However, it is often considered that techniques such as even-order Sum- or Difference-Frequency-Generation (SFG and DFG respectively) cannot be used to probe the bulk of centrosymmetric media.
Pioneering works by Bethune et al. \cite{bethune1976optical,bethune1977measurement, bethune1981optical,bethune1981quadrupole} have demonstrated that these considerations do not hold when quadrupolar interactions are considered.
By specifically targeting an electric quadrupole transition in one step of the interaction pathway, they were able to show that resonant second-order SFG (which is equivalent to resonant three-wave mixing: TWM) in centrosymmetric media has a quadrupolar origin.
Bethune et al. \cite{bethune1978sum}, Kim et al.\cite{kim1997second}, Okada et al. \cite{okada1981optical}  all have measured the quadrupolar resonant SFG signals in various bulk alkali metal vapor gases (Na, K, Li), relying on their high nonlinear hyperpolarizabilities \cite{fuentealba1993polarizabilities, reintjes2012nonlinear}.

However, latter experiments focusing on non-resonant SHG in Na vapor by Miyazaki et al. \cite{miyazaki1979spontaneous} and in Xe by Malcuit et al. \cite{malcuit1990anomalies} did not match the predictions of a signal with quadrupolar coupling.
A model based on static electric fields arising from charge separation following photoionization was proposed to explain a laser-induced symmetry breaking.
Finally, an alternative explanation based on an induced anisotropy following spatial variation of the densities of ground state atoms along the focused laser beam propagation direction was proposed by Freeman et al. \cite{freeman1981optical}.

In this Letter, we demonstrate that second-order SFG can be measured off-resonance in various atomic gases (He, Ar, Kr) and diatomic molecules (O$_2$, N$_2$), and rule out quadrupolar SFG and charge separation-induced static electric fields as possible mechanisms. 
This work generalizes previous observations to molecular gases but also to SFG instead of only SHG.

We first present experimental results demonstrating the observation of off-resonant SFG processes. 
Measurements of pressure and intensity scaling of the signal demonstrate its nature.
Polarization characterization rules out the quadrupolar origin of the signal.

\section{Experiment setup}

\begin{figure*}[h!]
    \centering
    \includegraphics[width=1\textwidth]{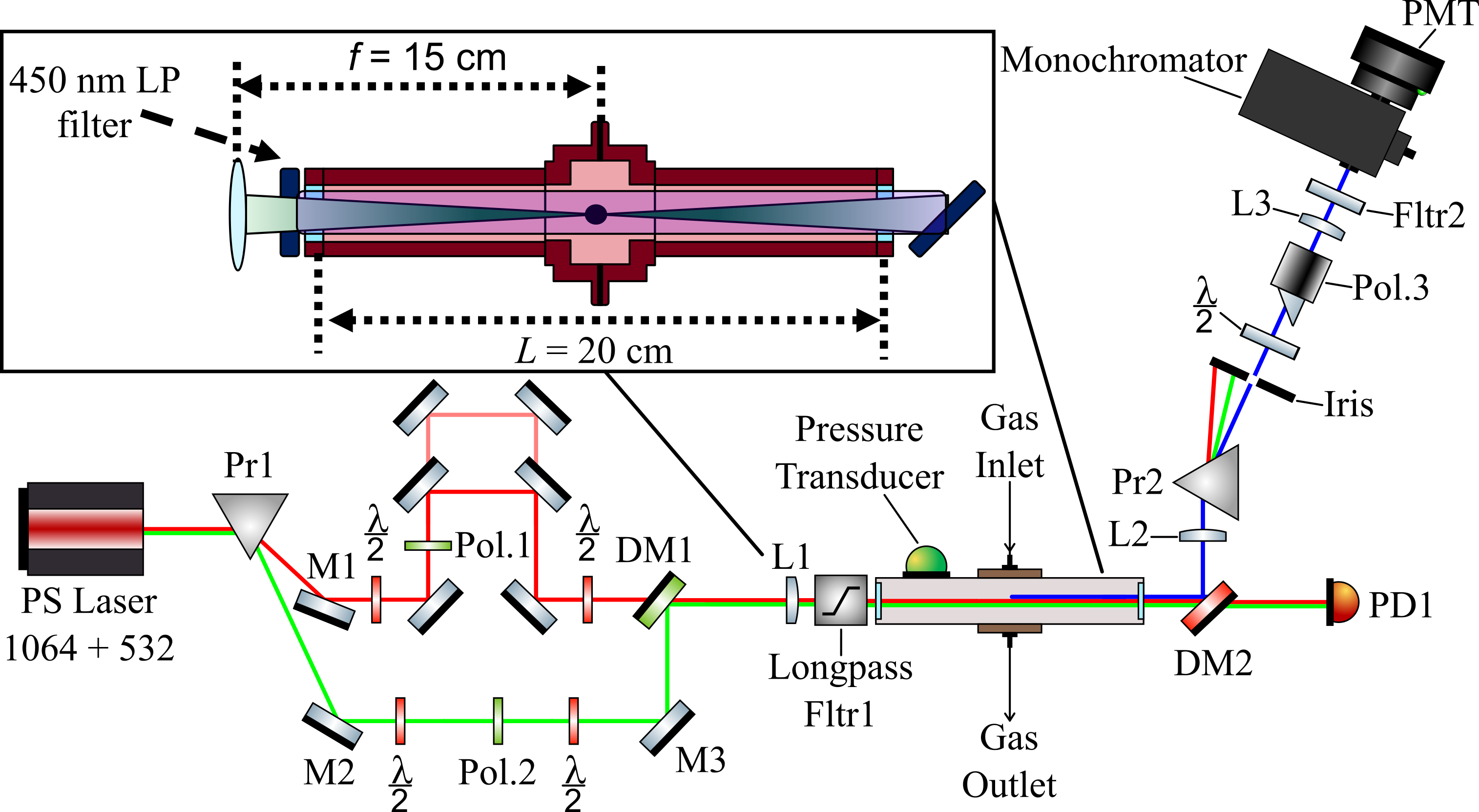}
    \caption{Schematic of the experimental setup.}
    \label{fig:setup}
\end{figure*} 

The experimental setup used to measure the SFG signal is displayed in Fig. \ref{fig:setup}.
The experiments are carried out using a 50Hz 1064 nm Nd:YAG laser system, which delivers 30mJ 30ps pulses.
A frequency-doubling crystal is used to generate a 532 nm second harmonic.
An equilateral dispersion prism spatially splits the two beams. 
Two half-waveplates and a polarizer are incorporated in each of the beam paths to allow independent control of the polarization and pulse energy. 
An optical delay line is used to obtain coincident pulses at the sample location.
The beam paths are recombined by a dichroic mirror (DM1) and focused with a $f=15$ cm N-BK7 AR-coated lens into a small pressure chamber.
Unless otherwise specified, the experiments are conducted using about $\mathcal E_1 = 1$ mJ at 1064 nm and about $\mathcal E_2 =0.1$ mJ at 532 nm. Therefore, the corresponding laser intensities at the focus are $I_1 = 3.2 \times 10^{12}$ W/cm$^2$ and $I_2 =1.2\times 10^{12}$ W/cm$^2$, respectively.
The pressure chamber is equipped with inlet and outlet Swagelok ports as well as a vacuum pump (Specstar, 9.6 cfm) and valves allowing to fill up the chamber with different high-purity gases. 
A baraton capacitance manometer (MKS, 628B13TDE1B) is used to monitor the gas pressure within the chamber. 
A dichroic mirror (DM2) transmits the 1064 and 532 nm beams while reflecting 355 nm. The transmitted beams are measured using a silicon photodiode (PD1) to monitor the intensities of the incident beams as well as for timing purposes.
The SFG signal is then collimated with a $f=15$ cm fused-silica plano-convex lens (L2), dispersed (Pr2) to remove any remnant of the 1064 and 532 beams and focused (L3, $f=10$ cm) onto a photomultiplier tube (PMT, Hamamatsu, C5594-44).
The 355 nm signal beam is further isolated before the PMT using a narrow bandpass filter (FLH355-10, Center 355 nm, FWHM 10 nm) and a manual Mini-Chrom monochromator (Edmund Optics, Model C, slit 300 $\mu$m, 1800 grooves/mm, resolution 2.18 nm).
The polarization state of the SFG signal is measured using  a halfwave plate and a polarizer (Pol.3) prior to the PMT. 

The absence of any artifact leading to Third Harmonic Generation (THG) of the fundamental beam or SFG from the window chamber was checked as follows.
A 450 nm hard-coated long-pass filter (OD>5) is inserted in between the focusing lens and the test chamber entrance window to remove any stray 355 nm signal produced by any optics ahead of the chamber. 
The entrance window is made of N-BK7 material, with a specified transmission of less than 10\% below 500 nm. 
The incident beams are about 4 mm in diameter at the entrance window of the test vessel, strongly limiting nonlinear interactions arising from that window. 
The incident beams are focused near the center of the pressure chamber with a beam waist of $\sim$40 $\mu$m.
The laser-induced breakdown threshold for a typical 1064 nm 40 ps laser at 1 atm in Ar is about  10$^{13}$ W/cm$^2$ \cite{ireland1974gas, morgan1975laser}. 
The pulse intensities used in the setup exclude the possibility of the 355 nm signal generated through laser-induced plasma-related effects.

\section{Results and discussion}
\subsection{Laser intensities and gas pressure dependencies}

\begin{figure}[h!]
    \centering
    \includegraphics[width=0.48\textwidth]{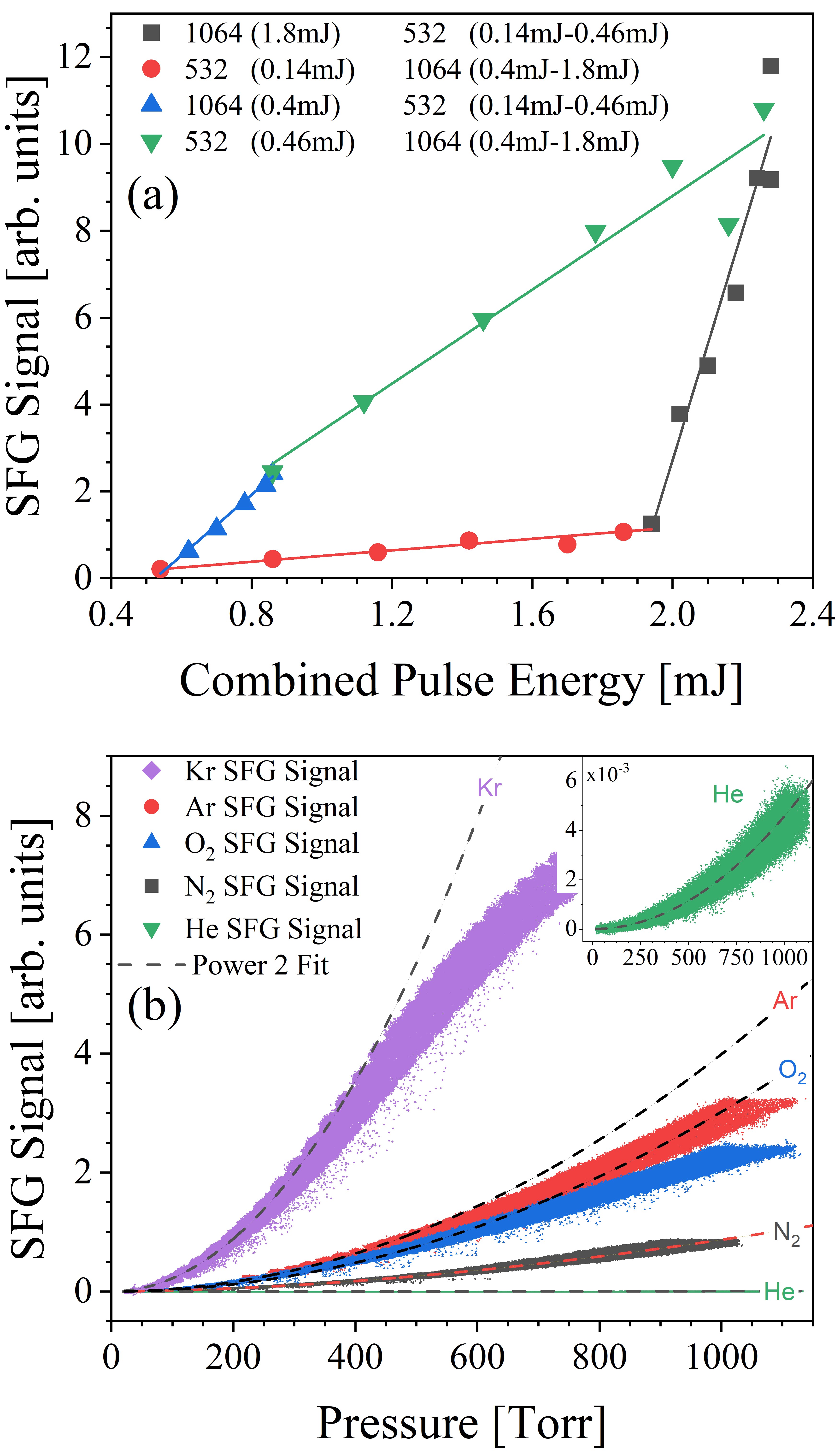}
    \caption{{(a) SFG signal in room air at 355 nm as a function of the combined power of the 1064 and 532 nm input beams, (b) Scaled intensity of the SFG signal at 355 nm for 5 different gases as a function of the gas pressure with a power of 2 fitting for each gas. }}
    \label{fig:powscaleb}
\end{figure}

\begin{table}[h!]
\begin{center}
    \caption{ Relative nonlinear hyperpolarizabilities of five atomic and molecular gases \cite{finn1971dc,shelton1994measurements}} 
    \begin{tabular}{c c}

    \hline
        Species & Nonlinear hyperpolarizability [arb. units]\\
    \hline
        He & 1 \\
        N$_2$ & 21 \\
        O$_2$ & 31 \\
        Ar & 32 \\
        Kr & 77 \\
    \hline
    
    \end{tabular}

    \label{table:S1}
\end{center}
\end{table}

Fig. \ref{fig:powscaleb} (a) displays the intensity scaling of SFG signals using air in the chamber with respect to the pulses energies $\mathcal E_1$ and $\mathcal E_2$. 
The measured signal increases linearly with both  $\mathcal E_1$ (black square and blue triangle symbols) and $\mathcal E_2$ (red circle and green downward triangle symbols). The latter observation is consistent with the expected intensity scaling for SFG given by
\begin{equation}
I_\text{SFG} \propto |\chi^{(2)}|^2 I_1 I_2
\label{eq:1}
\end{equation}
\noindent where $I_\text{SFG}$ is the intensity of the SFG signal, $\chi^{(2)}$ is the second-order nonlinear susceptibility. 
Furthermore, $\chi^{(2)}$ scales linearly with $n$, where $n$ is the number density of the gas molecules, and thus $I_\text{SFG}$ scales quadratically with $n$.
The contribution of THG to the signal measured at 355 nm is negligible (see supplemental material). 
The linear dependencies highlighted in Fig. \ref{fig:powscaleb} (a) are also observed in room air when the test chamber is removed from the beam path, ruling out possible artifacts generated by the chamber windows.

Figure \ref{fig:powscaleb} (b) displays the SFG signal strength as a function of the pressure for various gases considered. 
Vertically-polarized incident beams were used for these measurements.  
The measurements were performed between 20 and 1100 Torr in Kr, Ar, O$_2$, N$_2$, He. 
For all five investigated gases, the SFG signal increases with the gas pressure. Furthermore, we observe that \textit{I$_\text{SFG}$$^\text{(2)}$} (Kr) $\gg$ \textit{I$_\text{355}$$^\text{(2)}$} (Ar) $\cong$ \textit{I$_\text{SFG}$$^\text{(2)}$} (O$_2$) > \textit{I$_\text{SFG}$$^\text{(2)}$} (N$_2$) $\gg$ \textit{I$_\text{SFG}$$^\text{(2)}$} (He). 
These results are in good agreement with the relative magnitudes of the investigated gases hyperpolarizations \cite{finn1971dc, shelton1994measurements}, see Table 1. 
The signal scalings in figure \ref{fig:powscaleb} (b) can be all fitted with quadratic functions in low pressure ranges, as expected by a quadratic scaling in the number density $n$ of gas particles.
Departures from the quadratic scaling are observed for larger pressures and correspond to signal amplitudes in which the PMT's response is nonlinear. 
Departures from perfect quadratic functions are most visible for gases with large nonlinear hyperpolarizabilities.
The scaling on the number density observed in Fig. \ref{fig:powscaleb} (b) suggests the possibility of leveraging non-resonant three-wave mixing to measure number densities at the picosecond timescale in virtually all gases. We note that the same $n^2$ dependency was also reported by Miyazaki et al. for the case of non-resonant optical second harmonic generation in Na vapor \cite{miyazaki1979spontaneous} whereas a $n^3$ dependency was observed by Malcuit et al in Xe for a 1064 nm laser intensity of 4×10$^1$$^3$ W/cm$^2$ (an order of magnitude larger) \cite{malcuit1990anomalies}.

\begin{figure}[h!]
    \centering
    \includegraphics[width=0.48\textwidth]{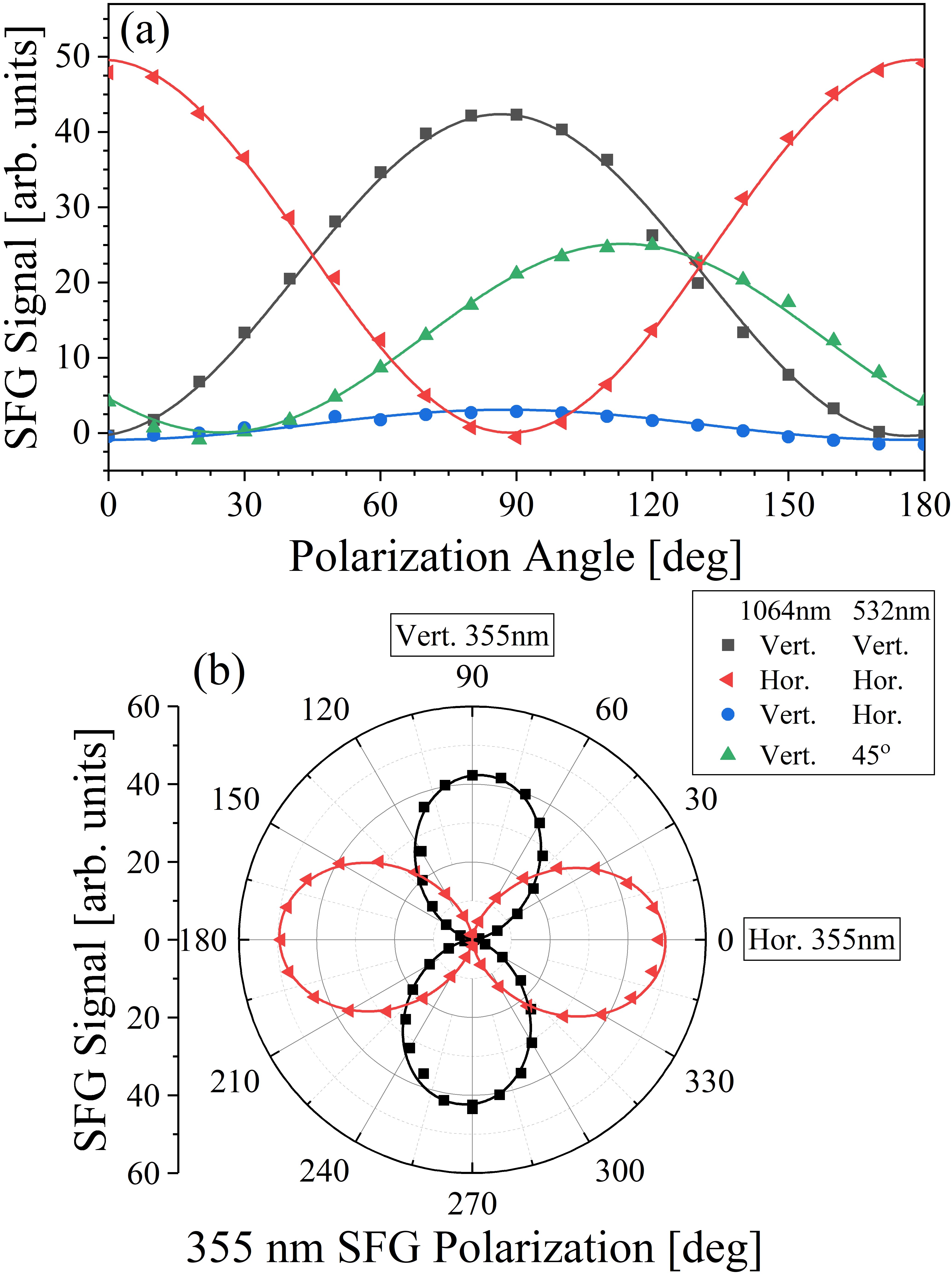}
    \label{fig:polarplot}
    \caption{{ (a) Polarization state of the TWM signal at 355 nm in room air. A sinusoidal wave is fitted for each combination. (b) Polar plot representing the polarization of the 355 nm signal for V1064-V532 and for H1064-H532 combinations.}
    \label{fig:polarplot}}
\end{figure}

\subsection{SFG Polarization}
 
We now discuss the polarization of the measured $I_\text{SFG}$ signal  as a function of the incident beam polarizations. 
In Figs. \ref{fig:polarplot} (a)-(b), it can be observed that when both incident beams have the same polarization, the emitted SFG has an identical polarization. 
For example, for vertically-polarized incident beams (V-V: black square), a vertically-polarized SFG signal is obtained. 
Likewise, for horizontally-polarized incident beams (H-H: red triangle), a horizontally-polarized SFG signal is measured.  
These observations are incompatible with both experimental observations and theoretical predictions of resonant quadrupolar coupling origin for the SFG signal \cite{bethune1978sum, flusberg1977optical}. 
A quadrupolar SFG signal is only expected when the incident laser beams are in a noncollinear phase matching arrangement, with orthogonal polarizations components. 

When incoming crossed polarization states are used, the polarization state of the SFG signal is different. For a vertically-polarized $I_1$ and a horizontally-polarized $I_2$, the SFG signal is polarized vertically (see Fig. \ref{fig:polarplot} (a) ). 
For a vertically-polarized $I_1$ and a 45$^\circ$ polarized $I_2$, the SFG is linearly polarized at $\sim$120$^\circ$. 
Additionally, it is observed that some SFG signal can be generated using orthogonally-polarized incident beams. However, the intensity of the mixed output signal in the latter case is significantly lower than that of the collinearly-polarized beams configuration. In general, our results demonstrates that the polarization of the non-resonant SFG signal depends on the polarizations of both incident beams and suggest that electric quadrupole effects are not responsible for our observations. 
These observations also differs from Miyazaki et al. \cite{miyazaki1979spontaneous, miyazaki1981interaction} showcasing no dependence of the intensity of the SHG signal on the incident beam polarization angle for non-resonant SHG. 
The latter authors ascribed their results to laser-induced multiphoton ionization resulting in charge separation and therefore static electric field generation (on the order of 50 V/cm for 10$^{12}$ W/cm$^2$ 28 ps laser pulses).

To test this hypothesis, we performed experiments in room air with externally applied electric fields up to 20 kV/cm (see supplemental material). 
No change in the measured SFG signal was observed, suggesting a different mechanism than proposed by Miyazaki et al. 
Freeman et al. \cite{freeman1981optical} suggested the observed effects could be qualitatively interpreted as a result of the alteration of isotropic properties of the probed gases following a spatial variation of the ground state densities of the probed atoms/molecules.
With the latter spatial density variations due to the combined effects of spatial variation of the laser beam intensity and multiphoton absorption.  Although we are not yet able to pinpoint the details of the mechanism responsible for all the evidenced nonlinear effects, all these observations make the case for the suitability of leveraging non-resonant picosecond three-wave mixing for number density measurements in both flows and plasma environments.

\section{Conclusions}

In this Letter, we have presented experimental observations of non-resonant optical SFG in various atomic and molecular gases.
This nonlinear effect was previously believed to be forbidden in the gas phase.
Across all investigated gases, a quadratic relationship between the SFG signal intensity and the ambient gas pressure is observed. Moreover, the signal is independent of any externally applied E-Field, indicating its potential for gas number density measurements on the ps timescale for multiple cases including in low and high-pressure reactive flow environments and in plasmas.

\section*{Acknowledgements}

We thank Dr. Donald Bethune and Prof. Yuen Ron Shen for insightful discussions.
We acknowledge the financial support of the College of Science and Engineering and the Department of Mechanical Engineering of the University of Minnesota, Twin Cities.
Work by JRR was supported by the U.S. Department of Energy  (DOE), Office of Science, Basic Energy Science  (BES), Chemical Sciences, Geosciences and Biosciences Division (CSGB) under Contract No. DE-AC02-06CH11357.

\bibliographystyle{unsrt}
\bibliography{biblio}

\end{document}